\documentstyle[preprint,aps]{revtex}
\begin{document}
\draft
\tighten
\hyphenation{ana-logue ana-lo-gous ana-lo-gous-ly ana-ly-sis}

\title{Locality and Causality in Hidden Variables Models of 
Quantum Theory}

\author{Stefan Teufel, Karin Berndl, Detlef D\"urr}
\address{Mathematisches Institut der Universit\"{a}t 
M\"{u}nchen,\\
Theresienstra{\ss}e 39,
80333 M\"{u}nchen, Germany}
\author{Sheldon Goldstein}
\address{Department of Mathematics,
Rutgers University,\\ New Brunswick, NJ 08903, USA}
\author{Nino Zangh\`{\i}} 
\address{Istituto di Fisica dell'Universit\`a di Genova,\\ 
Istituto Nazionale di Fisica Nucleare, Sezione di Genova,\\ 
Via Dodecaneso 33, 16146 Genova, Italy} 
\date{April 30, 1997}
\maketitle

\begin{abstract}
Motivated by Popescu's example of 
hidden nonlocality, we elaborate on the
conjecture that quantum states that 
are intuitively nonlocal, i.e.,
entangled, do not admit a local causal 
hidden variables model. We exhibit
quantum states which either (i) are 
nontrivial counterexamples to this
conjecture or (ii) possess a new kind 
of more deeply hidden irreducible
nonlocality. Moreover, we propose a 
nonlocality complexity classification
scheme suggested by the latter possibility. 
Furthermore, we show that
Werner's (and similar) hidden variables 
models can be extended to an
important class of generalized 
observables.  Finally a result of Fine on
the equivalence of stochastic and 
deterministic hidden variables is
generalized to causal models.

\end{abstract}


\narrowtext

\section{Introduction}

The nonlocality inherent in the quantum mechanical 
description of nature is 
still a much debated subject, 30 years after the 
pioneering work by John Bell 
\cite{Bell}. Bell's starting point was the argument 
of Einstein, Podolsky, 
and Rosen (EPR) \cite{EPR}, 
demonstrating the existence of local hidden 
variables implying the incompleteness of the quantum 
mechanical description of a physical system 
by the wave function alone. The EPR argument is 
based on one essential 
assumption: {\it locality, or local causality.} 
Bell derived an inequality 
which the distributions of these local hidden 
variables have to satisfy
and showed that the quantum mechanical values 
for these probabilities violate 
this inequality. Thus there cannot be a local 
hidden variables model for quantum 
theory. But Bell's argument in conjunction with 
the EPR argument actually 
shows much more: the quantum mechanical predictions 
cannot be explained by 
local physical laws. And since the quantum mechanical 
predictions are confirmed 
by most experiments, see for example \cite{Aspect}, one 
has to conclude that 
there are nonlocal physical actions in nature.

EPR argued by considering a thought experiment with a certain
correlated 2-particle state. Bohm transformed this thought 
experiment into 
one which is similar to actually performed experiments, 
involving spin measurements 
on the singlet state \cite{BohmEPR}. Does this EPR-Bell 
nonlocality arise only in 
such special situations, while most quantum systems are 
local? For most quantum 
states the EPR argument clearly does not apply since it 
requires perfect 
correlation between lots of pairs of results of widely 
separated measurements. 
Bell has introduced an alternative to the EPR-argument 
that applies to all quantum 
states: the concept of stochastic local hidden variables 
\cite{Bell}.
As has been noted (in a special case) by Fine \cite{Fine}, 
this concept actually 
leads to the same framework as the EPR argument, 
namely to (deterministic) 
local hidden variables. 

Thus one may now ask which quantum states admit a 
local hidden variables model, 
i.e., are EPR-Bell-local. For pure quantum states 
one can rather easily see that 
for any entangled state of 2 or more particles one 
can find observables whose 
correlations violate a Bell inequality 
(see for example  \cite{Gisin,PopRohr}). This 
implies that no local 
hidden variables model exists for such quantum 
states. Only pure product states 
$\psi = \psi^{(1)} 
\otimes \dots \otimes \psi^{(N)}$ are EPR-Bell-local. 
Now what do we expect for 
mixed states? A mixed 2-particle state describing 
an ensemble 
of pure product states, i.e, a convex sum of 
product density matrices $\rho = 
\sum _\nu p_\nu \rho_{\nu}^{(1)} \otimes \rho_{\nu}^{(2)}$, 
$p_\nu > 0 $, is also 
EPR-Bell-local. The question is whether any ``entangled'' 
density matrix, i.e., 
any density matrix that cannot be written as a mixture of 
product density matrices, 
is EPR-Bell-nonlocal. We certainly expect that such states 
exhibit nonlocal behavior because
in any representation of the density matrix as a convex 
sum of pure states there are entangled
wave functions present. The problem is how to reveal 
this nonlocality.

That this question is tricky is suggested 
by the existence of
a class of density matrices discovered by 
Werner which are entangled but nevertheless 
admit a local hidden variables model \cite{Werner}.
The Werner states $W$ are states on the Hilbert space 
${\cal H} = 
{\ {\sf l}\!\!\!{\rm C}}^{d} \otimes {\ {\sf l}\!\!\!
{\rm C}}^{d}$ 
and may be thought of as describing  two spin $({d-1})/{2}$ 
particles:
\begin{equation}\label{W} 
W=  \frac{d+1}{d^3}\,  I - \frac 1{d^2} \, F ,
\end{equation}
where $F$ is the flip operator, $F (\psi^{(1)} \otimes 
\psi^{(2)}) = \psi^{(2)} 
\otimes \psi^{(1)}$. Recently Popescu \cite{Popescu} 
pointed out an interesting 
property of Werner states. 
When we perform local measurements 
of the form 
$P= T \otimes I$ and $Q = I \otimes T$ 
on each subsystem, where $T$ is a 
projector on a two-dimensional 
subspace of 
${\ {\sf l}\!\!\!{\rm C}}^d $, $T=|1\rangle \langle 1| + 
|2\rangle \langle 2|$, 
we get nonlocal correlations in a sub-ensemble. The 
sub-ensemble for which $P$ 
and $Q$ both yield the outcome $1$, described by the 
collapsed density matrix 
\begin{equation} \label{W'}
 W' = \frac{(P Q )W(P Q)}{{\mbox{tr}\, } (WPQ)},
\end{equation} 
violates a Bell inequality for $d \geq 5$; thus 
there are local
observables whose correlations in this new state can 
no longer be described
by a local model. The nonlocality of $W$---intuitively 
corresponding to the
property of $W$ of being entangled, but ``hidden'' when 
only a single
measurement on each side is considered---can thus be 
revealed by {\it
sequences}\/ of local measurements.\footnote{Gisin 
\cite{Gisin2} has
recently found entangled states in $d=2$  which do not 
violate certain types of Bell
inequalities for single measurements but do so for 
sequences of (generalized) measurements.}

Since in Popescu's example the observables leading 
to a violation of a Bell
inequality commute with $P$ and $Q$, these time 
sequences of measurements
can be described by single observables, and Werner's 
model may be applied
to it. However, the model thus obtained violates 
causality: later
measurements influence preceding ones 
\cite{Popescu,MerminBiel}. 
Since such
a model does not appropriately represent 
the physical idea of
locality---actions into the past are even worse than 
actions at a
distance---we propose, following Popescu \cite{Popescu} 
and Mermin
\cite{MerminBiel}, as a requirement for a local hidden 
variables model that
it satisfy the locality and causality conditions, 
defining an LCHV model,
described here.\footnote{Recently, the causality 
condition for stochastic
hidden variables models has also been discussed by 
Zukowski et al.\
\cite{ZHHH}.}

 This is discussed in Section \ref{lchvsect}: in 
subsection \ref{detlchvsect}
for the deterministic case, and in subsection 
\ref{stochlchvsect} for the stochastic 
case. It turns out however that any quantum 
state that allows
for a stochastic hidden variables model 
also allows for a deterministic model
and vice versa. This 
is a  generalization of a result of Fine 
(to causal models with arbitrary many 
observables).

In Section \ref{conjsect} we shall 
discuss our main conjecture about the
classification of local states---that 
for $d>2$ no entangled density matrix
on ${\ {\sf l}\!\!\!{\rm C}}^{d} 
\otimes {\ {\sf l}\!\!\!{\rm C}}^{d}$
admits an LCHV model. We can neither 
prove nor disprove this conjecture,
but we shall discuss some partial results 
which shed light on it.  For pure
product states and mixtures of product 
states there exists, as one would
certainly expect, an LCHV model. This 
is shown in Section
\ref{productsect}. Likewise, for pure 
entangled states there can be no LCHV
model. In Section \ref{popescusect} we 
discuss Popescu's result in our
terms: the Werner states (for $d \geq 5$) 
do not admit LCHV models. In Section
\ref{countersect} we show that in 
$d = 2$ dimensions the conjecture is
false. All the same, the case $d=2$ 
is special, and the conjecture may well
be correct for $d\geq 3$. We exhibit a 
larger class of states similar to
the Werner states, which are entangled, 
admit a local hidden variables
model, and have the additional property 
that after the first measurement on
one side the projected states are 
mixtures of product states, i.e., local
states.  However, in contrast to the 
2-dimensional case we cannot conclude
that there is an LCHV model for these states.  

This situation is interesting: Either 
these states have an LCHV model and
hence are counterexamples to our main 
conjecture, or they do not, in which
case they possess a more deeply hidden 
irreducible nonlocality, different
from that of the examples known so far, 
and suggesting a new nonlocality
classification scheme, described in 
Section \ref{classect}.

In Section \ref{povsect} we shall consider 
generalized measurements: since
Werner's local hidden variables model is a 
model for standard observables only, 
it may be
possible, as has been conjectured by Popescu 
\cite{Popescu}, to reveal the 
``hidden nonlocality'' by generalized 
observables associated with positive 
operator valued (POV) measures. More 
generally, the existence of a local hidden 
variables model for generalized observables 
might be equivalent to the state 
being non-entangled. 
However, we show in Section \ref{extensect} 
that local hidden variables models 
for standard observables can be extended 
to cover some POV measurements, which 
restricts the class of POV's that can be used to
prove or disprove this conjecture.

\section{Local causal hidden variables 
models}\label{lchvsect}

\subsection{Deterministic hidden 
variables}\label{detlchvsect}

To motivate our definition of hidden 
variables models we shall briefly
repeat the essentials of the EPR 
argument in Bohm's version. Consider two
spin-$\frac{1}{2}$ particles in the 
singlet state. If the spin of each
particle is measured in some direction 
$\bf a$ the two outcomes are
perfectly anticorrelated, but the 
outcome of a particular measurement is
completely random. If one particle 
has $\bf a$-spin $+1$ the measurement of
the $\bf a$-spin of the other particle 
will yield $-1$.  The assumption of
locality now implies that the particles 
carry the information about their
spin value in every direction with them. 
To account for the perfect
anticorrelation as well as for the 
randomness of the particular result, the
adequate mathematical language is that 
of random variables $X_{{\bf a}},
X_{{\bf b}}$ for the spin in every 
direction $\bf a$ and $\bf b$,
respectively, for both particles on a 
probability space $\Omega$ with a
probability measure ${\rm I\! P}$.\footnote{The 
following analogy may be 
helpful. In classical statistical mechanics 
the microstate of a system is a 
point in phase space and the statistical 
properties of macroscopic systems 
are explained by considering macrostates, 
i.e., ensembles of systems in
different microstates which are macroscopically
indistinguishable. Observables are just 
random variables on phase space and
their distributions are determined by the 
macrostate. Can the quantum
mechanical observables be understood in a 
similar way? Systems with the
same wave function would, in this analogy, 
belong to the same macrostate
and the individual properties of a particular 
system would be determined by
the microstate, a point in a yet unknown 
probability space, traditionally
called the space of hidden variables. 
Quantum mechanical observables would
be random variables on this new probability 
space and their distributions
would agree with the quantum mechanical 
predictions for suitably chosen
ensembles of microstates.  (In particular, 
there would be joint
distributions for {\it all}\/ observables.)}  
We shall sometimes call this
kind of hidden variables ``deterministic 
hidden variables''---the hidden
variable $\omega\in\Omega$ determines the 
results of all possible
experiments---to distinguish them 
from the ``stochastic hidden
variables'' we shall introduce in the 
next subsection. Note, however, that
determinism has been inferred from locality.
 
We shall consider the following general setup. 
The physical system is in a
quantum state $\rho$, which is a density matrix 
on a Hilbert space $\cal
H$. For the time being, we shall assume that only
ideal (von Neumann) measurements are performed
on the system, i.e., a measurement is associated
with an observable, which is a self-adjoint 
operator on $\cal H$, and the usual collapse
rule for sequences of ideal measurements applies.
Observables are denoted
by $O^k$, their spectral values by $o$, 
and their spectrum by $\sigma
(O^k)$. Furthermore, we denote by $P^k_o$ 
the projection onto the
eigenspace of the observable $O^k$ corresponding 
to the eigenvalue $o$. For
simplicity we shall explicitly focus only on 
observables with discrete
spectrum.

By a {\bf hidden variables model (HV model)} 
for a quantum state $\rho$ 
we shall mean  a probability space $(\Omega , 
{\rm I\! P} )$
and a family  $X$ of  random variables on 
$(\Omega , {\rm I\! P} )$ such
that  for 
any time-ordered sequence of (Heisenberg) observables
$(O^1 (t_1),\ldots ,O^n (t_n))$ with $t_1 < 
\ldots < t_n$ of arbitrary 
length $n$ there is a corresponding random 
variable, a function
\[ X_{O^1 (t_1),\ldots,O^n (t_n)}: \Omega \to 
\sigma(O^1 ) \times \ldots   
\times \sigma(O^n ) \subset {\rm I\! R}^{n} , \]
whose distribution agrees with the quantum 
mechanical distribution 
for the results of a  sequence
of ideal  measurements 
of the observables $O^1 (t_1),\ldots ,O^n (t_n)$:
\begin{eqnarray} & & {\rm I\! P} \left( X_{O^1 
(t_1) ,\ldots,O^n (t_n)}=(o_{1} ,\ldots,
o_{n})\right) \nonumber\\ & & \quad =
{\mbox{tr}\, } \left( P^n_{o_{n}} (t_n) \cdots 
P^1_{o_{1}} (t_1 ) \, \rho \,  
P^{1}_{o_{1}}(t_1 ) \cdots P^{n}_{o_{n}}(t_n ) 
\right) .\label{qprob}\end{eqnarray}

This condition guarantees 
that the HV model captures all the quantum 
mechanical predictions for
arbitrary (ideal) 
measurements in the state $\rho$. (If 
in the sequence $(O^1 (t_1),\ldots ,O^n (t_n))$
some consecutive observables are jointly
measurable, i.e., commute, we may allow the 
times of the consecutive commuting 
observables to be equal.)

For any quantum state $\rho$, there is a 
HV model: for example, take $(\Omega , 
{\rm I\! P} )$ to be the interval $[0,1]$ 
with uniform distribution (=Lebesgue measure), 
and define the random variables by splitting 
$[0,1]$ into pieces with lengths equal to the 
probabilities of the occurring values. Note 
that this trivial construction does not 
contradict the no-hidden-variables theorems. 
The proofs that there can be no hidden 
variables require  additional properties of 
the hidden variables. One such
property is that of ``non-contextuality;'' 
see \cite{Bell,nrao} for a 
discussion of this property and its limited 
physical relevance. We shall now consider 
physically relevant additional properties, 
namely locality and causality. 

 We split the random variable  $ X_{O^1(t_1) ,\ldots ,
O^n(t_n)} $ that describes a sequence of $n$ measurements
into an initial segment $X^{O^n(t_n)}_{O^1(t_1), 
\ldots ,O^{n-1}
(t_{n-1})}$ of length $n-1$ describing the 
outcome of the first $n-1$ measurements
which in general depends on the later 
measurement $O^n(t_n)$, 
and a final (``follow-up'') 
random variable  $X_{O^n(t_n) }
^{O^1(t_1),\ldots,O^{n-1}(t_{n-1})}$, describing the 
outcome of the last experiment $O^n(t_n)$ 
depending on the first $n-1$ measurements: 
\begin{eqnarray*} & &  X_{O^1(t_1) ,\ldots ,
O^n(t_n)} \\ & & \quad =: 
\left( X^{O^n(t_n)}_{O^1(t_1), \ldots ,O^{n-1}
(t_{n-1})}, X_{O^n(t_n) }
^{O^1(t_1),\ldots,O^{n-1}(t_{n-1})} \right). 
\end{eqnarray*}

A {\bf causal 
hidden variables model (CHV model)} for a 
quantum state $\rho$ is a HV model 
satisfying the causality condition that 
initial segments do not depend upon what
measurements are later performed, i.e., that 
\begin{equation}\label{cc}
X^{O^n(t_n)}_{O^1(t_1),...,O^{n-1}(t_{n-1})}=
X_{O^1(t_1),...,O^{n-1}(t_{n-1})}.
\end{equation}
Equivalently, the HV model is  causal if
\begin{eqnarray} & & X_{O^1(t_1) ,\ldots ,
O^n(t_n)} \nonumber\\ & & \quad = 
\left( X_{O^1(t_1)}, X_{O^2(t_2)}^{O^1(t_1)}, 
\dots , X_{O^n(t_n) }^
{O^1(t_1),\ldots,O^{n-1}(t_{n-1})} \right) .
\label{caus}\end{eqnarray}
This condition captures the physical idea of 
causality: the description of an 
experiment associated with $O^m(t_m)$  may 
depend on those experiments performed 
earlier, but it is not changed by experiments 
that are performed in the future.
If, for example, two sequences of observables 
are considered differing in the later 
observable, $(O^1(t_1), O^2(t_2))$, $t_1<t_2$, 
and $(O^1(t_1), {O^2}'(t_2'))$, 
$t_1<t_2'$, then 
\[ X_{O^1(t_1) ,O^2(t_2)} = \left( X_{O^1(t_1)}, 
X_{O^2(t_2)}^{O^1(t_1)} \right) \]
and 
\[ X_{O^1(t_1) ,{O^2}'(t_2')} = \left( X_{O^1(t_1)}, 
X_{{O^{2}}'(t_2')}^{O^1(t_1)} \right) ,\]
and the outcome of $O^1(t_1)$
does not depend upon which, if any, 
measurements are performed later.

Just as for HV models, one can readily 
see, since the quantum probabilities
(\ref{qprob}) are, in an obvious sense, 
causal, that all quantum mechanical
states admit a CHV model. Take $(\Omega 
, {\rm I\! P} )$ to be $[0,1]$ with uniform
distribution, define the one-observable 
random variables $X_{O^1(t_1)}$ for
all $O^1(t_1)$ as explained above, and 
define then inductively on each of
the subsets $\{ X_{O^1(t_1), \ldots,O^{n-1}
(t_{n-1})} = (o_1,\dots ,
o_{n-1}) \}$ the follow-up random 
variables $ X_{O^n(t_n)
}^{O^1(t_1),\ldots ,O^{n-1}(t_{n-1})}$ 
by further splitting according to
the conditional probabilities of the 
occurring values \cite{Diplom}.

To discuss locality, we consider the 
following generalized EPR
situation. The system is in a quantum 
state $\rho$ on the Hilbert space
${\cal H} = {\cal H}^{(1)} \otimes 
{\cal H}^{(2)}$, the factors of which
correspond to two subsystems that are 
spatially separated such that no
signal traveling at most at the speed 
of light will be able to propagate
between them while two observers are 
locally conducting their experiments.
Local measurements are described by 
observables of the form $A = A^{(1)}
\otimes I$, on system 1, and $B = I 
\otimes B^{(2)}$, on system 2. As a
matter of convenience, when considering 
a sequence of local measurements we
shall rearrange the random variables 
describing the successive
measurements, placing those for system 
2 to the right of those for system 1
and writing $X_{A^1,\dots , A^n,B^1,\dots , 
B^m}$ accordingly, regardless
of their relative time ordering.\footnote{To 
simplify the notation
we will sometimes drop the time variable of 
the Heisenberg operators.}
 (Note that since in this setup all the
$A$'s commute with all the $B$'s, we may 
collect together the $A$'s and the
$B$'s in formula (\ref{qprob}) as well.)

A {\bf local hidden variables model (LHV model)} is a 
HV model satisfying the locality  condition that
the random variables describing the outcomes of 
local measurements on one system 
do not depend on the measurements performed 
on the other system, i.e., 
\[ X_{A^1,\dots , A^n,B^1,\dots , B^m} = 
( X_{A^1,\dots , A^n}, X_{B^1,\dots , 
B^m} ) .\]
This condition says that while the results 
of local measurements in the two 
separated subsystems may be correlated, 
as for example are the colors of Bertlmann's 
socks \cite{Bell}, the local measurements 
performed on one system do not 
themselves influence those on the other system.

For the case of one local observable 
for each subsystem, this condition
reduces to the condition that we have 
random variables $X_A$ and $X_B$
for all observables $A$ and $B$ for 
each subsystem---so that $X_{A,B} =
(X_A, X_B)$---such that the pair 
distributions agree with the quantum
mechanical ones. This is the traditional 
framework of local hidden
variables. The causality condition is
obsolete in this case. We shall abbreviate by 
{\bf LHV1 model} such
an LHV model for single measurements, 
which we shall often speak of as
measurements at a single time.

Finally we define a {\bf local causal 
hidden variables model (LCHV model)}
as a HV model that is both local and causal;
and an LCHV$k$ model as a model for sequences 
of measurements of length $k$ on both sides. 
(LCHV1 = LHV1.)

\subsection{Stochastic hidden 
variables}\label{stochlchvsect} 

Bell has introduced also so-called 
``stochastic local hidden variables models,'' 
where the hidden variable doesn't determine 
the measurement results completely, 
so that given the hidden variable 
the results are still random. The role of the 
hidden variables is to {\it explain}\/ 
correlations between measurement results at 
distant places: Assuming locality, it must 
be possible to identify sufficiently many 
causal factors=hidden variables, such that 
the residual fluctuations of the outcomes of 
experiments will be independent if these 
causal factors=hidden variables are held 
fixed (see Bell \cite{Bell}, pp.\ 150). 

A {\bf stochastic local causal HV model} 
for a quantum state $\rho$ consists of a 
probability space $( \widetilde{\Omega} , 
\widetilde{{\rm I\! P}} )$ and a family of 
maps $ Q_{O^1 (t_1) ,\ldots ,O^n (t_n) }$: 
for any time-ordered sequence 
of arbitrary length $n$ of observables $(O^1 
(t_1) ,\ldots ,O^n (t_n))$ with $t_1 < 
\ldots < t_n$  there is a map
from $\widetilde{\Omega}$ into probability 
distributions on the product of the spectra
$ \sigma(O^1 ) \times \ldots \times 
\sigma(O^n )$ denoted by
\begin{equation} Q_{O^1 (t_1) ,\ldots ,
O^n (t_n) }: \widetilde \omega
\mapsto   Q_{O^1 (t_1) ,\ldots ,O^n (t_n) } 
(\cdot\,,\dots ,\cdot\,| \, 
\widetilde\omega ) \end{equation} with the 
following properties: (a) By
averaging over $\widetilde{\Omega}$ the 
quantum mechanical probabilities are obtained
\begin{eqnarray} & &  \int_{\widetilde{\Omega}} 
Q_{O^1 (t_1) ,\ldots ,O^n (t_n) }( o_{1} ,
\ldots,o_{n} | \, \widetilde \omega ) \, 
d\widetilde{{\rm I\! P}} ( \widetilde \omega) 
\nonumber\\ & & \quad = {\mbox{tr}\, } \left( 
P^n_{o_{n}}(t_n) \cdots P^1_{o_{1}}(t_1) \, 
\rho\,  P^{1}_{o_{1}}(t_1) 
\cdots P^{n}_{o_{n}}(t_n) \right) .
\label{QProb}\end{eqnarray}
(b) Causality:  The marginals do not depend on 
later measurements
\begin{eqnarray*} & & Q_{O^1(t_1) ,\ldots ,
O^n(t_n)}( o_{1} ,\ldots,o_{{n-1}} | \, 
\widetilde \omega ) \\ & & \quad \equiv  
\displaystyle \sum_{o_{{n}} \in \sigma (O^n)} 
Q_{O^1(t_1) ,\ldots ,O^n(t_n)}( o_{1} ,
\ldots,o_{{n-1}}, o_{{n}}| \, \widetilde 
\omega ) \\ & & \quad = \ Q_{O^1(t_1) ,
\ldots ,O^{n-1}(t_{n-1})}( o_{1} ,\ldots,o_{{n-1}}| 
\, \widetilde \omega ); \end{eqnarray*}
equivalently
\begin{eqnarray} & & Q_{O^1(t_1) ,\ldots ,O^n(t_n)}
( o_{1} ,\ldots,o_{{n}}| \, 
\widetilde \omega ) \nonumber\\ & & \quad  =  
Q_{O^1(t_1) ,\ldots ,O^{n-1}(t_{n-1})}
( o_{1} ,\ldots,
o_{{n-1}}| \, \widetilde \omega ) \nonumber\\ 
& & \quad \quad \cdot \, Q_{O^1(t_1) ,\ldots ,
O^{n}(t_{n})}( o_{n} | \, o_{1} ,
\ldots,o_{{n-1}} , \widetilde \omega ) 
\nonumber\\
& &  \quad =  Q_{O^1(t_1)}( o_{1} | \, \widetilde 
\omega ) \, Q_{O^1(t_1) ,O^{2}(t_{2})}
( o_{2} | \, o_{1} , \widetilde \omega ) \nonumber\\
& &  \quad \quad \,\cdots \,  Q_{O^1(t_1) ,\ldots ,O^{n}
(t_{n})}( o_{n} | \, o_{1} ,\ldots,o_{{n-1}}, 
\widetilde \omega ) \label{bedingt}
\end{eqnarray}
with the conditional probabilities
\begin{eqnarray*} &  &  Q_{O^1 (t_1) ,\ldots ,
O^n (t_n) }( o_{{k+1}}, \dots , o_{n}| o_{1} ,
\ldots,o_{k} , \, \widetilde \omega ) \\
& & \quad = \frac {Q_{O^1 (t_1) ,\ldots ,O^n
(t_n) }( o_{1} ,\ldots,o_{k} , o_{{k+1}}, \dots , 
o_{n}| \, \widetilde
\omega )}{Q_{O^1 (t_1) ,\ldots ,O^k (t_k) }
( o_{1}, \dots , o_{k}| \,
\widetilde \omega )} .\end{eqnarray*} 
(c) Locality: In the same sort of EPR framework as
described earlier, the separated systems are 
conditionally independent
given $ \widetilde \omega $, i.e., for local 
measurements the probabilities
$Q_{\dots} (\dots | \widetilde \omega )$ 
for fixed $ \widetilde \omega $
factorize, 
\begin{eqnarray*} & &  Q_{A^1,\ldots,
A^{n},B^1,\dots , B^m}(a_{1} ,
\ldots,a_{{n}},b_{1} ,\ldots,b_{{m}}|\,
\widetilde \omega) \\
& & \quad = Q_{A^1,\ldots,A^{n} }(a_{1} ,
\ldots,a_{{n}}|\,\widetilde \omega) Q_{B^1,
\dots , B^m}( b_{1} ,\ldots,b_{{m}}|\,
\widetilde \omega),\end{eqnarray*} 
where we have used the same rearrangement as earlier.
Equivalently,
\begin{eqnarray*}  & Q_{A^1,\ldots,A^{n},
B^1,\dots , B^m} ( a_{1} ,
\ldots,a_{{n}}| \, b_{1} ,\ldots,b_{{m}} , 
\widetilde \omega ) & \\ & =
Q_{A^1,\ldots,A^{n},B^1,\dots , B^m} ( a_{1},
\ldots,a_{{n}}| \, \widetilde
\omega ) & \\ & = Q_{A^{1}, \dots ,A^{n} } 
( a_{1} ,\ldots,a_{{n}}| \,
\widetilde \omega ) & \end{eqnarray*} and 
analogously for the other system.
(The first equality directly above is 
usually called ``outcome
independence'' and the second one 
``parameter independence.'') In the same spirit we 
may define stochastic 
HV, CHV, LHV1, LCHV, and LCHV$k$ models.
 
The existence of a deterministic model for a 
quantum state trivially implies the existence 
of a stochastic model with the same locality 
and causality properties.
Interestingly, the converse is also true: From a 
stochastic model for a quantum state one may 
construct a deterministic model with the same 
locality and causality properties.
Fine has proven the equivalence between 
the existence of deterministic and
stochastic LHV1 models \cite{Fine}. While 
he explicitly considers models
involving only 4 observables, the basic 
idea of his proof extends to an
arbitrary number, and in fact to all LHV 
models as well. We show in Appendix \ref{equisect} 
that the equivalence between the existence of 
deterministic and stochastic
models holds also for LCHV models.

This result is interesting for several reasons.
 Firstly, it reduces the complexity of a classification 
of states concerning nonlocality, since any state 
allowing for a stochastic model of a certain kind 
automatically also allows for a deterministic model 
with the same properties and vice versa.
Secondly, when proving the existence of a HV model 
for a certain quantum state,
we may use Werner's construction of a stochastic HV 
model and conclude
that there is also a deterministic model. We shall 
use similar reasoning in the following 
discussion several times.
Thirdly, we wish to remark that the equivalence between 
the existence of deterministic and stochastic HV models 
is also conceptually interesting with respect to the 
discussion of physical nonlocality. The EPR argument 
shows that locality 
implies the existence of deterministic LHV, while Bell's 
argument, as explained at the beginning of Section 
\ref{stochlchvsect}, derives the concept of stochastic 
LHV from locality. These two seemingly different 
approaches are actually equivalent.

\section{Is the nonexistence of an LCHV model 
equivalent to the 
state being entangled?}\label{conjsect}

\subsection{The case of pure states and non-entangled 
density matrices}\label{productsect}

Pure entangled states do not admit a local hidden 
variables model 
\cite{Bell,Gisin,PopRohr}. Thus there certainly 
can be no LCHV model for such states. 

For quantum states which are a mixture of product 
density matrices one can construct an 
LCHV model by appropriately ``mixing'' the LCHV 
models for the individual 
terms.  For completeness,
we briefly give the natural construction. 
Consider for simplicity a 
finite sum $\rho = \sum_{\nu=1}^{n} p_{\nu} 
\rho_{\nu}^{(1)} \otimes 
\rho_{\nu}^{(2)}$ with  $p_{\nu} 
\geq 0 $ for all $\nu$, $\sum_{\nu=1}^{n} 
p_{\nu} =1$, and $(\Omega^{(1)}, 
{\rm I\! P} ^{(1)}, X^{(1,\nu )})$, $(\Omega^{(2)}, 
{\rm I\! P}^{(2)} , X^{(2,\nu )})$ 
CHV models for $\rho^{(1)}_{\nu}$ and  
$\rho^{(2)}_{\nu}$. (For all $\nu$, 
the probability space $(\Omega^{(i)}, 
{\rm I\! P} ^{(i)})$ may be chosen to be 
$[0,1]$ with Lebesgue measure.)
Define $ \Omega := 
\Omega^{(1)} \times \Omega^{(2)} \times 
\{ 1,\dots ,n \} $
and $
{\rm I\! P} := {\rm I\! P} ^{(1)} \times 
{\rm I\! P}^{(2)} \times p  ,$ where  
$p(\{\nu\})=p_\nu$ for ${\nu}\in\{ 1,\dots,n \}$.
For sequences of local observables 
$A^i = A^{(1)i} \otimes I$, $B^j = I 
\otimes B^{(2)j} $ let 
\begin{eqnarray*} & &  X_{A^1,\ldots ,
A^n, B^1,\dots , B^m } (\omega  ) \\
& & \quad := \left( X^{(1,\nu )}_{A^{(1)1},
\ldots ,A^{(1)n}} 
(\omega^{(1)} ), X^{(2,\nu )}_{B^{(2)1},
\dots , B^{(2)m} } (\omega^{(2)}) \right) , 
\end{eqnarray*}
$\omega = (\omega ^{(1)}, \omega ^{(2)}, \nu) $.  
One readily sees that this is an LCHV 
model for $\rho$.
Thus all non-entangled density matrices 
are local in the sense that they admit an 
LCHV model. (This of course covers non-entangled 
pure states.) 

By a 
construction similar to the one
just given, one sees that the set of 
states (in some fixed Hilbert space)
admitting an LCHV model, like the set of 
non-entangled states, is convex,
i.e., if $\rho_1$ and $\rho_2$ admit an 
LCHV model, then so does $t\rho_1
+(1-t)\rho_2$ for $0<t<1$.

In section C below we exhibit for $d=2$ an 
entangled density matrix that
nevertheless admits an LCHV model. However, 
as will be seen in section C, the 
case $d=2$ is rather special, so
that we still find it reasonable to 
conjecture that {\it for $d>2$ no
entangled density matrix admits an LCHV model.}  
We can neither prove nor
disprove this conjecture. Recently, 
necessary and sufficient criteria for
the quantum state to be entangled have 
been found \cite{Peres,Horodecki}.
However, we are unable to use these 
criteria in our framework.  In the
following we shall present some partial results.

\subsection{Popescu's result: there is no 
LCHV model for the 
Werner states for $d\geq 5$}\label{popescusect}

 A CHV model for a quantum state $\rho$ 
naturally yields 
one for collapsed states arising from 
$\rho$. Consider a measurement of $A$, and 
suppose the outcome 
is $a$. The collapsed quantum state is 
$\rho '= \left( P_{a} \rho P_{a}\right) 
/ {\mbox{tr}\, } \left( \rho \, P_{a} 
\right)$. {From} a CHV model $( \Omega ,
{\rm I\! P} , X)$ for 
$\rho$ one constructs in the obvious way 
one for $\rho '$: Let 
\[ \Omega ' := \{ X_{A} = a \} ,  \ 
{\rm I\! P} ' = \frac{{\rm I\! P} |_{\Omega '}}
{{\rm I\! P} (\Omega ')} \]
and put on $\Omega '$
\begin{eqnarray*} & & X'_{A^1, \dots , 
A^n} := X_{A^1, \dots , A^n}^A \\
& & \quad := \left( X_{A^1}^A, 
X_{A^2}^{A, A^1}, 
\dots , X_{A^n}^{A, A^1, \dots , A^{n-1}} 
\right) \end{eqnarray*} 
This model gives the correct quantum 
probabilities (\ref{qprob}) and
satisfies the causality condition 
(\ref{cc}). Moreover,
\[ X _{A^n}^{'A^1, \dots , A^{n-1}} = 
X_{A^n}^{A, A^1, \dots , A^{n-1}} .\]

Furthermore, if $( \Omega ,{\rm I\! P} , X)$ 
is a {\it local}\/ CHV model for
$\rho$, then $( \Omega ' ,{\rm I\! P} ' , X')$ 
is a local CHV model for the
collapsed state $\rho '$. In our framework, 
Popescu's result takes the
following form: There is no LCHV model for 
the Werner states $W$ (Eq.\
(\ref{W})) for $d\geq 5$ since there is no such 
model for the collapsed
state $W'$ (Eq.\ (\ref{W'})) for $d\geq 5$.

We also wish to note the absolutely crucial 
role played by causality here:
Regardless of whether or not it is local, if 
$( \Omega ,{\rm I\! P} , X)$ were not
causal then $( \Omega ' ,{\rm I\! P} ' , X')$ 
need not be well defined, since 
$\Omega'$ could then depend upon the 
later measurements.
 
\subsection{A counterexample in $d=2$ and 
``almost counterexamples'' in $d \geq 3$}
\label{countersect}

A possible way to come to grips with our 
conjecture is to find a counterexample.
We shall now discuss a family of generalized 
Werner states (compare
Eq. (\ref{W}))
\begin{equation} \label{WDef}
W = \frac{1}{d} \left( \frac{1}{d} +
 c \right) I - c\,F
\end{equation}
which may turn out to provide counterexamples.  
If the real parameter $c$
varies between $0$ and $\frac{1}{d^2 -d}$ 
these states are normalized (${\mbox{tr}\, }
W=1$) and positive.  Werner has shown that 
the states (\ref{WDef}) are
entangled if and only if ${\mbox{tr}\, } 
(F W) \, < \, 0$, which translates to $c \, >
\, \frac{1}{d(d^2 -1)}$.  For $c = 
\frac{1}{d^2}$ we recover the original
Werner states, for which Werner has 
constructed a stochastic LHV1 model. 
Since for $c \leq \frac{1}{d^2}$, 
$W$ is a mixture of the original Werner 
state and a multiple of the identity, 
which clearly allows 
for an LHV1 model,  by the mixing construction 
of section \ref{productsect} all $W$ with $c 
\leq \frac{1}{d^2}$ allow for an LHV1 model.

For the $d = 2$ Werner state, Popescu has already 
remarked that his nonlocality argument  does not 
apply \cite{Popescu}. We will now in fact construct an 
LCHV model for the state $W$ with
$c\leq \frac{1}{4}$ with the help of the 
LHV1 model just described. In a
two-dimensional Hilbert space  any 
nontrivial, i.e.,
$\neq \mbox{const.}I$, observable is 
nondegenerate, so that its projectors
are onto one-dimensional
subspaces. Therefore after the measurement 
of the first two local
observables\footnote{We may assume 
without loss of generality that
nontrivial measurements are performed on 
each subsystem.} $A^1$ and $B^1$ the 
collapsed state is a pure product state,
\begin{eqnarray*} W' & = & \frac{ (P^{(1)}_{A^{1}
={a}} \otimes P^{(2)}_{B^{1}={b}}) W 
(P^{(1)}_{A^{1}={a}} \otimes P^{(2)}_{B^{1}={b}})}
{{\mbox{tr}\, } (P^{(1)}_{A^{1}={a}} \otimes 
P^{(2)}_{B^{1}={b}}) W} \\
& = & P^{(1)}_{A^{1}={a}} 
\otimes P^{(2)}_{B^{1}={b}}
, \end{eqnarray*} 
and hence has an LCHV model. Moreover, 
the states $ P^{(1)}_{A^{1}={a}}$ and
$P^{(2)}_{B^{1}={b}}$ describing the 
subsystems after the measurements {\em do not}
depend on what measurement has been performed 
on the other subsystem. {\em Therefore}
given an LHV1 model $(\Omega^{(W)},{\rm I\! 
P}^{(W)}, X^{(W)})$ for the state $W$
and, for all pure states $\alpha= 
P^{(1)}_{A^{1}={a}}$ and $\beta=P^{(2)}_{B^{1}={b}}$, 
CHV models $(\Omega^{(1)}, {\rm I\! P}^{(1)}, 
X^{(1,\alpha)})$ and $(\Omega^{(2)},
{\rm I\! P}^{(2)}, X^{(2,\beta)})$ (where, 
without loss of generality, we have
assumed that the probability spaces 
$(\Omega^{(1)}, {\rm I\! P}^{(1)})$ and
$(\Omega^{(2)}, {\rm I\! P}^{(2)})$ don't 
depend on $\alpha$ and $\beta$,
respectively) one can obtain an LCHV model by 
a simple coupling
of the LHV1 model with the CHV models:

Let
\[ \Omega := \Omega^{(W)} \times \Omega^{(1)}
 \times \Omega^{(2)}, \ \ {\rm I\! P}  := 
{\rm I\! P} ^{(W)} \times {\rm I\! P}^{(1)} 
\times {\rm I\! P}^{(2)}  \]
and
\begin{eqnarray*}  X_{A^1,\ldots,A^n} 
( \omega) & := & \left( X^{(W)}_{A^1} 
(\omega^{(W)}), X^{A^1}_{A^2,\ldots,A^n} 
(\omega^{(W)},\omega^{(1)}) \right), \\ 
   X_{B^1,\ldots,B^m} ( \omega) & := & 
\left( X^{(W)}_{B^1} (\omega^{(W)}), 
X^{B^1}_{B^2,\ldots,B^m} (\omega^{(W)},
\omega^{(2)}) \right), \end{eqnarray*}
with 
\[  X^{A^1}_{A^2,\ldots,A^n} (\omega^{(W)},
\omega^{(1)}) := X^{(1,\alpha_{A^1}
(\omega^{(W)}))}_{A^{2(1)},\ldots,A^{n(1)}} 
(\omega^{(1)})\]
where
\[\alpha_{A^1}(\omega^{(W)})=P^{(1)}_{A^{1}
=X^{(W)}_{A^1}(\omega^{(W)})}
,\]
and
\[  X^{B^1}_{B^2,\ldots,B^m} (\omega^{(W)},
\omega^{(2)}) := X^{(2, \beta_{B^1}
(\omega^{(W)}))}_{B^{2(2)},\ldots,B^{m(2)}} 
(\omega^{(2)}) \]
where 
\[\beta_{B^1}(\omega^{(W)})=P^{(2)}_{B^{1} 
= X^{(W)}_{B^1}(\omega^{(W)})} .\]
It is easy to see that $( \Omega, {\rm I\! P},X)$
defines an LCHV model for $W$. Thus we have 
found a family of entangled states, 
namely the generalized Werner states (\ref{WDef})  
with $\frac 16 < c \leq \frac 14$  
in a $2\times 2$-dimensional Hilbert space, 
which nevertheless admit an LCHV model. We shall 
call these states  $W_{d=2}$ states.

We shall next show that in dimension $d\geq 3$
the (entangled) Werner states with $c \in 
\left( \frac{1}{d(d^2 -1)},  \frac{1}
{d(d^2 -1) - d^2} \right] $ become non-entangled
already after one nontrivial local measurement 
(on one side). (This is trivially true
in $d=2$.) We shall later 
refer to these states
as $W_{d\geq 3}$ states.
To show that they are left in a non-entangled state 
after one 
nontrivial local measurement, note that if $P$ 
is a projector on a $(d-1)$-dimensional  
subspace of 
${\ {\sf l}\!\!\!{\rm C}}^d$, 
the collapsed state $W' = (P\otimes P)\, W(P\otimes P) / 
{\mbox{tr}\, } (W(P\otimes P))$  on the  
subspace $\text{Ran} P \otimes 
\text{Ran} P$ has again the form (\ref{WDef}) 
with \[ c' = \frac{cd^2}{(d-1)(d-cd-1)}. \]
For $ c' \leq \frac{1}{(d-1)( (d-1)^2
 -1)} $ $\Leftrightarrow$ 
$ c \leq \frac{1}{d(d^2 -1) - d^2}$ the state 
$W'$ is non-entangled, and from this 
it follows that the same
thing is true when the range of $P$ has 
dimension $< (d-1)$.  Thus we calculate
\begin{eqnarray} & & (P \otimes I)\, W 
(P \otimes I) \nonumber\\ & & \quad
= (P \otimes (P + P^{\perp}))\, W (P 
\otimes (P + P^{\perp})) \nonumber\\ &
& \quad = (P \otimes P)\, W (P \otimes 
P) \nonumber\\ & & \quad \quad +
\frac 1d \left( \frac 1d +c\right) (P 
\otimes P^{ \perp})\, I (P \otimes
P^{\perp}) \nonumber\\ & & \quad \quad 
-c (P \otimes P^{ \perp})\, F (P
\otimes P^{\perp}) \nonumber\\ & & 
\quad \quad - c (P \otimes P^{ \perp})\,
F (P \otimes P)\nonumber\\ & & \quad 
\quad -c (P \otimes P)\, F (P \otimes
P^{\perp}).  \label{sum}\end{eqnarray} 
Here $P^{\perp} = I-P$.  The first
term in the sum (\ref{sum}) is non-entangled 
due to the foregoing argument,
the second term is obviously non-entangled, 
and a straightforward
calculation shows that the last three 
terms vanish.  Thus already after one
local measurement the collapsed state is 
non-entangled and therefore local.

Thus for any $d\geq 2$ we have found 
states that are entangled, admit an
LHV1 model, and are left after any 
local measurement in a non-entangled
state, where there is of course an 
LCHV model for further measurements.
However, we cannot extend our construction 
of an LCHV model for the original
state from $d=2$ to $d\geq 3$, since 
the range of the projectors may have
dimension $>1$, in which case the state 
after one local measurement on each
side may depend irreducibly on both 
measurements, i.e., in such a manner
that random variables on one side must 
have some dependence on what
observable was first measured on the
other side: The LHV1 model for the
state itself contains random variables 
$X_{A^1}$ and $X_{B^1}$ for the
first measurements.  The existence of 
an LCHV model for all collapsed states
yields the follow-up random variables 
$X^{A^1 , B^1}_{A^2 , \ldots , A^n}$
and $X^{A^1 , B^1}_{B^2, \ldots , B^m}$, 
not the follow-up random variables
$X^{A^1}_{A^2 , \ldots , A^n}$ and 
$X^{B^1}_{B^2, \ldots , B^m}$ required
for an LCHV model for the original state. 
Nevertheless, it does follow that
it is impossible to reveal the nonlocality 
of these states by a
Popescu-type argument, namely by 
producing a nonlocal state violating a
Bell inequality by local measurements.

\subsection{A Nonlocality Classification Scheme}
\label{classect}

We have shown that for $d=2$ the main 
conjecture is wrong, i.e.\ there are 
entangled mixed states allowing for an LCHV model. 
For the  $W_{d \geq 3}$ entangled mixed 
states allowing for an LHV1 model for the 
first measurement and an LCHV model for 
all further measurements, both possibilities, 
either the existence of an LCHV model or its 
nonexistence, are very interesting.

In the first case the main conjecture 
is indeed wrong: entangledness is not
equivalent to the nonexistence of an 
LCHV model.  If, on the other hand, it
turns out that there is no LCHV model 
for these states, a new nonlocality
complexity class emerges: a class of 
entangled mixed states with
nonlocality that is more deeply hidden 
than that of the Popescu/Werner
example.

We are thus lead to a nonlocality 
classification scheme which we shall
briefly describe.  We classify the 
nonlocality of quantum states by means
of a pair $(N,n)$ of positive integers, 
the {\it indices of nonlocality.\/}
Larger values of these indices 
correspond to more deeply hidden
nonlocality.  The first index $N$ 
corresponds to the length of the sequence
of measurements necessary to reveal the 
nonlocality, and the second index
$n$ conveys the degree to which the 
state can be transformed, by performing
local measurements upon the system, 
to one in which the nonlocality is more
manifest. More precisely, the first 
index denotes the smallest integer $N$
such that there is no LCHV$N$ model 
(i.e., no local causal model for
sequences of maximal length $N$) for 
the states in this class. For example,
entangled pure states have $N = 1$ while 
Werner's states for $d \geq 5$
have $N = 2$. For states that have an 
LCHV model we put $N =
\infty$.\footnote{Abstract considerations 
show that the existence of an
LCHV$k$ model for all finite $k$ 
implies the existence of an LCHV model.}

Suppose a state has finite first index 
of nonlocality $N$. After one
measurement the state collapses into one 
that clearly has first index at
least $N-1$.  We say ``at least'' 
because it may well be the case that the
collapsed state has an LCHV$k$ model 
for $k\geq N-1$, i.e., has index $N$
or greater. For example, for our $W_{d \geq 3}$ 
states, the state after one
nontrivial measurement has index $N=\infty$. 
 In contrast, in Popescu's
example the index $N=2$ can be reduced to 
$N=1$ by one measurement.  One
may thus hope to reduce the first index of 
nonlocality by performing a
sequence of measurements.  The smallest 
number to which it can, with
nonvanishing probability, be reduced 
will be denoted by $n$. Note that if
there is no LCHV model for a $W_{d\geq 3}$ 
state, the nonlocality of this
state must be {\it irreducible\/}, 
i.e., we must have that $n=N$.

$N=\infty$ for nonentangled states, 
which are completely local. But there
are also entangled states allowing 
for an LCHV model, as is the case, for
example, with the $W_{d=2}$ states.  
It would seem natural to conclude that
such states do not produce any nonlocal 
effects.  However, a recent result
by Bennett et al.\ \cite{Purif} shows 
that this conclusion is wrong.  They
found a nonlocality argument employing 
a sequence of local unitary
operations and measurements on an 
ensemble of systems each in such a
state---the unitary operations and 
measurements are performed on states of
the form $W \otimes W \otimes 
\cdots \otimes W$ on the Hilbert 
space ${\
{\sf l}\!\!\!{\rm C}}^4 \otimes {\ 
{\sf l}\!\!\!{\rm C}}^4 \otimes \cdots
\otimes {\ {\sf l}\!\!\!{\rm C}}^4$. 
 These states possess nonlocality that
is even more deeply hidden than what 
we have discussed so far. This
suggests now as the second classification 
index $n$ in the case $N=\infty$
the minimal number of copies of the 
system in the ensemble required for
such a nonlocality argument. If there 
is no finite $n$ with this property,
we put $n=\infty$. 

The following table contains the 
known results about
this classification:

\bigskip

\begin{tabular}{|c|c|}
\hline
$(N,n)$ & examples \\
\hline 
 $(1,1)$ & entangled pure states\\
 $(2,1)$ & Werner states for $d\geq 5$ \\
 $(k,k)$ for some $k\geq 2$ & $W_{d\geq 3}$ 
states having no LCHV model\\
 $(\infty , k)$ with $k<\infty$ & $W_{d=2}$ states\\
$(\infty , ?)$ & $W_{d\geq 3}$ states which 
have an LCHV model \\
$(\infty , \infty )$ & nonentangled states\\
\hline
\end{tabular}

\bigskip

\noindent A natural conjecture---a weakening of 
our main conjecture---is
that the only states in class $(\infty , \infty )$ 
are the nonentangled ones.

A further possibility to exhibit 
nonlocality is the use of generalized 
measurements, which we shall discuss below. 
It may turn out that analogous classification 
indices are of interest also for generalized 
measurements, but we have no results in that 
direction.

\section{Generalized observables}\label{povsect}
\subsection{Hidden variables models for 
generalized observables}
 
We shall now extend the allowed class of
 observables.
In standard quantum theory the measurement 
of an observable is associated
with a self-adjoint operator, which 
corresponds by the spectral theorem to
a projection-valued (PV) measure.  But the 
concept of PV measure can be
generalized in a natural way to that of 
positive operator valued (POV)
measure.  POV's have been
employed to construct generalized quantum 
observables for which the usual
framework of self-adjoint operators has 
been unsuccessful, mainly in the fields
of quantum optics and the theory of 
open quantum systems
\cite{Davies}. Moreover, POV's emerge 
naturally from an analysis of
quantum experiments in Bohmian mechanics 
\cite{DDGZ}.  A POV measure is a
set function $M$, which maps measurable 
subsets $\Delta\subset {\rm I\! R}$ to
bounded positive operators $M({\Delta})$, 
such that for any quantum state
$\rho$
\[ \Delta \mapsto \mu^{\rho}_M (\Delta) 
:= {\mbox{tr}\, } \rho\, M({\Delta}) \]
is an ordinary probability measure on 
${\rm I\! R}$, i.e., $M({\rm I\! R})=I$ 
and for any sequence 
of disjoint measurable sets $(\Delta_i)_
{i\in {\rm I\! N}}$, $\Delta_i\subset 
{\rm I\! R}$, 
\[ M\Bigl( {\bigcup_{i\in {\rm I\! N}} 
\Delta_i}\Bigr) = \sum_{i\in {\rm I\! N}} 
M({\Delta_i}).\] 
$\mu^{\rho}_M (\Delta)$ is the probability 
in the quantum state $\rho$ for
the outcome of the measurement associated
 with the POV $M$ to be in
$\Delta$. The complete formal description 
of a measurement of a generalized
observable, whose outcome statistics are
 given by a POV $M$, is provided by
the so-called operations $R_{\delta}$, 
a family of bounded linear
operators, describing the change of quantum 
state during the measurement if
the outcome $\delta\in {\rm I\! R}$ is obtained:
\[ \rho \rightarrow \tilde \rho = \frac{R_{\delta} 
\rho R^{\dagger}_{\delta}}
{{\mbox{tr}\, }  \rho\, R_\delta ^\dagger R_\delta }.\]
The operations $R_\delta$ determine 
the POV $M$ via 
\[ M(\Delta ) = \sum_{\delta\in\Delta} 
R^\dagger_\delta R_\delta .\] 
POV measures include as a special case 
the PV measures where all the
positive operators are orthogonal 
projections. (In an ``ideal'' measurement
of a PV $M$ the operations $R_\delta$ are 
equal to the projections
$P_\delta =M(\delta) := M(\{ \delta\} )$ 
corresponding to the usual
collapse rule. However, in general there 
are huge classes of possible
operations $R_\delta$ giving rise to the 
same POV or PV via $M(
\delta ) = R_\delta^\dagger R_\delta$.)

As with ordinary observables, we consider
 here only generalized observables
having a discrete set of possible values 
$\delta$, and we denote by $\cal R$
the corresponding family of operations 
$\{R_{\delta}\}$.

Our definitions of the various kinds of 
hidden variables models may without 
effort be generalized  to cover
measurements of generalized observables: 
just consider, instead of sequences 
of observables $O^1(t_1),\dots , 
O^n(t_n)$, sequences ${\cal R}^1(t_1),
\dots , {\cal R}^n(t_n)$. 
$\sigma({\cal R}^k)$ is the generalized 
spectrum of the generalized measurement 
${\cal R}^k$, i.e., the set of possible 
outcomes $\{ \delta\ |\  R^k_\delta\neq0\}$. 
Eq. (\ref{qprob}) for the quantum mechanical 
probabilities gets replaced by
\begin{eqnarray} & & 
 {\rm I\! P} \left( X_{{\cal R}^1 (t_1) ,
\ldots,{\cal R}^n (t_n)}=(\delta _{1} ,\ldots,
\delta _{n})\right) \nonumber\\
& & \quad = {\mbox{tr}\, } \left( R^n_{\delta 
_{n}} (t_n) \cdots R^1_{\delta_{1}} (t_1 )
\, \rho \, R^{1\dagger }_{\delta_{1}}(t_1 ) 
\cdots R^{n\dagger}_{\delta_{n}}(t_n )
\right) .\label{povprob}
\end{eqnarray}
The causality and locality condition do 
not change at all, and similarly
the definition of a stochastic local 
causal hidden variables model may be
trivially extended to cover measurements 
of generalized observables. 
We shall denote models which 
apply to generalized measurements 
by a G at the end of 
their abbreviation.

The equivalence theorem mentioned in 
Section \ref{lchvsect}, 
as well as its proof in Appendix \ref{equisect},
holds just as well for models for generalized 
observables, as do the
results in Sections \ref{productsect} and 
\ref{popescusect}. However, our
$d=2$ counterexample in Section \ref{countersect} 
to the conjecture that
only non-entangled states admit LCHV models 
does not generalize to LCHVG
models, since Werner's model is only for standard 
observables,\footnote{Actually only the 
first two measurements
(one on each side) have to be ideal measurements; 
the following ones may also
be generalized measurements.} as well as
because the state resulting from the 
first local measurements of 
generalized observables on the two sides 
may depend irreducibly on both
measurements, just as discussed in the 
second to the last paragraph of Section 
\ref{countersect}.  Thus the conjecture 
for generalized observables,
i.e., the conjecture that {\it only 
non-entangled states admit LCHVG models}, may be 
true in all dimensions $d\geq 2$.

However, in this context it would appear 
natural to first analyze a simpler 
conjecture which has been 
raised by Popescu \cite{Popescu}: {\it 
Is the existence of an LHV1G model equivalent to 
the state being non-entangled?\/}  
Is it possible to reveal the  nonlocality 
of any entangled state by
considering measurements at only a single 
time, at least if  generalized
observables are taken into account? Here 
again the 
Werner states serve as the first check: 
can Werner's model be extended 
to cover also generalized 
measurements? A positive answer to this 
question would answer Popescu's conjecture 
to the negative. In the following section 
we provide a partial positive answer.

\subsection{Extension of LHV1 models for 
standard observables to certain 
generalized observables}\label{extensect}

The LHV1 model constructed by Werner 
\cite{Werner} for the states (\ref{W})
covers only measurements of ordinary 
observables. In this section we will 
show that any
LHV1 model for ideal measurements can be 
extended to a model for certain
special generalized observables, namely 
those given by a POV involving  only  commuting
operators $M( \delta )$,\footnote{Since 
LHV1 models concern only a
single time, the operations $R_\delta$ 
describing the change of state during
measurement are not relevant here.} which 
we shall call commuting POV's. 
This covers for example all two-valued
POV's: If $M(\delta_ 1 ) + M(\delta_ 2 ) 
= I$, then $M(\delta_ 1 ) = I - M(\delta_ 2 )$
commutes with $M(\delta_ 2 )$. Moreover, 
many of the standard examples of
POV's are commuting POV's.  For example, 
given  an
observable $O=\sum _i o_i P_{o_i}$ in 
its spectral representation 
, we may consider the ``smeared-out''
projections
\[ M( o_i  ) = \sum_j t_{ij} P_{o_j} ,\]
with $t_{ij}\geq 0$ and $\sum_i t_{ij}=1$, 
clearly a commuting POV. Here 
$t_{ij}$ is the probability of 
obtaining the result $o_i$ when the system 
is in an eigenstate corresponding 
to the eigenvalue $o_j$. 
Also, the model of Gisin \cite{Gisin2} for
 a filtering 
process is described by a  commuting POV. 

We will now construct from an LHV1 model on 
$(\Omega,{\rm I\! P})$ for 
ordinary observables  a
stochastic LHV1 model, also on $(\Omega,
{\rm I\! P})$, for all generalized
observables governed by commuting 
POV's. By the equivalence
result of Appendix \ref{equisect} we 
have simultaneously a deterministic
LHV1 model for these generalized observables.  
Thus generalized observables
described by commuting POV's cannot 
reveal the ``hidden nonlocality'' of
the Werner states with measurements 
at a single time.

Let $M(\alpha  )=M^{(1)}(\alpha  ) 
\otimes I$ and $ \widetilde
M(\beta  )= I \otimes \widetilde M^{(2)}
(\beta  )$ each be
commuting POV's on a finite dimensional 
Hilbert space. (Using the notion of spectral 
measures, the following calculation can 
also be done for Hilbert spaces of 
infinite dimension \cite{Diplom}.)
A joint measurement of 
the corresponding generalized
observables is described by the POV $M
(\alpha  )\widetilde M(\beta 
)$, and the quantum mechanical probability 
for obtaining outcomes $\alpha$
and $\beta$ in this measurement  in the 
state $\rho$ is thus
\[ {\mbox{tr}\, } \rho \left( M^{(1)}(\alpha  ) \otimes
\widetilde M^{(2)}(\beta ) \right) .\] 
These are the probabilities that must 
be recovered by a stochastic local hidden 
variables model.

The positive commuting operators $M^{(1)}(\alpha  )$ 
have a joint
spectral representation
$M^{(1)}(\alpha  )= \sum_j m_{\alpha}^j 
P^{(1)}_j$ with $0 \leq
m_\alpha^j \leq 1$, where all the $P^{(1)}_j$ 
are projections onto  one-dimensional orthogonal 
subspaces and are independent of $\alpha$. 
Similarly, 
 $\widetilde M^{(2)}(\beta  )= \sum_k
\widetilde m_\beta^k \widetilde 
P^{(2)}_k$ with $0 \leq
\widetilde m_\beta^k\leq1.$

Let $A= \sum_j j P^{(1)}_j$ and 
$B=\sum_k k \widetilde P^{(2)}_k$.  
 Since we have assumed the existence 
of an LHV1 model $(\Omega
, {\rm I\! P}, X)$, 
we have random variables $X_{A}$,
$X_{B}$ for which 
\[{\rm I\! P}(X_A = j, X_B = k) = 
{\mbox{tr}\, } \rho (P^{(1)}_j \otimes P^{(2)}_k).\]
 We define 
\[ Q_M (\alpha |\omega) :=  
 \sum_j m_\alpha^j \openone_{\{j\}}(X_A(\omega)) \]
and 
\[Q_{\widetilde M} (\beta |\omega) := 
 \sum_k \widetilde m_\beta^k \openone_{\{k\}}
(X_B(\omega)) .\]
Then
\begin{eqnarray}
\int_{\Omega} & & Q_M (\alpha |\omega) Q_{\widetilde M}
(\beta |\omega) \, d{\rm I\! P} (\omega) \nonumber\\  
& &  = \sum_{j,k} m_{\alpha}^j \widetilde 
m_\beta^k \int_{\Omega} \openone_{\{j\}}
(X_A(\omega))  \openone_{\{k\}}(X_B(\omega))
 \, d{\rm I\! P} (\omega) \nonumber\\
&  & = \sum_{j,k} m_{\alpha}^j \widetilde 
m_\beta^k \ {\rm I\! P}(X_A = j , X_B = k) 
\nonumber\\
& &  = \sum_{j,k} m_{\alpha}^j \widetilde 
m_\beta^k \ {\mbox{tr}\, } \rho \left( 
P^{(1)}_j \otimes P^{(2)}_k \right)\nonumber \\
& & ={\mbox{tr}\, } \rho \left( \sum_j 
m_\alpha^j P^{(1)}_j \otimes \sum_k 
\widetilde m_\beta^k \widetilde P^{(2)}_k 
\right) \nonumber\\
& & =  {\mbox{tr}\, } \rho M(\alpha  )
\widetilde M(\beta ).
\label{povqmprob}\end{eqnarray}

Moreover, since
\[  I^{(1)} = \sum_{\alpha} M(\alpha  ) = 
\sum_j \sum_{\alpha}m_\alpha^j P^{(1)}_j    ,\]
it follows that 
\[ \sum_{\alpha} m_{\alpha }^j =1 \] 
for all $j$. Therefore the probability distribution 
$Q_M(\, .\, |\omega)$ is properly 
normalized:
\begin{equation}\label{povnorm} 
\sum_{\alpha} Q_M(\alpha |\omega) =1, 
\end{equation}
and analogously for $Q_{\widetilde M}
(\, .\, |\omega)$.  Thus from an LHV1
model on $(\Omega,{\rm I\! P})$ for 
ordinary observables we have constructed a
stochastic LHV1 model, also on $(\Omega,
{\rm I\! P})$, for all generalized
observables governed by commuting 
POV's.\footnote{Notice that the simpler
definition $Q_M(\alpha|\omega ) =
 X_{M( \alpha )} (\omega )$,
$Q_{\widetilde M}(\beta|\omega ) = 
X_{\widetilde M( \beta )} (\omega)$ also
yields the quantum mechanical 
probabilities (\ref{povqmprob}), and it does
so regardless of whether or not $M$ and 
$\widetilde M$ are each commuting
POV's. Note however that the normalization 
$\sum_{\alpha} Q_M(\alpha
|\omega) = 1$ need not hold with this 
definition, since it does not follow
from $\sum_{\alpha} M( \alpha ) = I$ 
that $\sum_{\alpha} X_{M( \alpha
)}=1$, regardless of whether the $M( \alpha )$ 
fail to commute (recall von
Neumann's no-hidden-variables theorem 
\cite[page 4]{Bell}) or do indeed
commute (recall Gleason's theorem 
\cite[page 6]{Bell}). Note also (again 
recalling
Gleason's theorem) that $X_{M( \alpha )}=
X_{m_{\alpha}(A)}$ need not agree
with $m_{\alpha}^{X_A}$ even in our 
commuting case.}

In this connection, it is interesting 
to comment on a recent work of Gisin
\cite{Gisin2}, who in Section 3 of 
his paper gives an example of an
apparent violation of the Bell-CHSH 
inequality even for non-entangled
states for generalized observables 
described by commuting POV's.  However,
his point that a ``careless application 
of generalized quantum measurements
can violate Bell's inequality even for 
mixtures of product states'' is
misleading: the observed violation of 
the Bell-CHSH inequality is not in
fact due to the application of generalized 
quantum measurements, but rather
to a misapplication of the inequality 
itself---to conditional expectations
in which the conditioning depends upon 
the measurements under consideration
\cite{CommGis}.

\section{Conclusion}

We have analyzed some questions concerning 
the classification of local
states in quantum theory. There are two 
intuitive notions of ``locality''
for quantum states: (i) the state is 
non-entangled, i.e., a product state
or a mixture of product states, and (ii) 
the state admits a local and
causal hidden variables model.

For pure quantum states it is well-known 
that the non-entangledness is
 equivalent to the existence of an LHV1 
model. We have shown that this
 equivalence extends to LCHVG models.  
For mixed states, the equivalence
 between non-entangledness and the 
existence of an LCHV model (for $d>2$)
 is our main conjecture. We have shown 
that (i) $\Rightarrow$ (ii), and
 have verified that pure states and 
Werner states with $d\geq 5$ conform to
 our conjecture. Furthermore, we have 
described counterexamples in
 dimension $d=2$, namely the states 
$W_{d=2}$, and a class of
 ``almost-counterexamples'' in higher 
dimensions, the states $W_{d\geq
 3}$. These ``almost-counterexamples'' 
leave two possibilities: Either the
 main conjecture is wrong in any 
dimension, or there is a more deeply
 hidden ``irreducible nonlocality,'' 
in the sense explained in Subsection
 \ref{classect}. In connection with 
the latter possibility we have proposed
two new indices of nonlocality as the 
basis of a nonlocality complexity
classification.  

Another conjecture concerning the 
classification of local states, which has
been raised by Popescu \cite{Popescu}, is 
that non-entangled states are the
only states admitting LHV1G models for 
{\it all}\/ possible measurements,
including those described by POV's. {From} 
our result in Section
\ref{povsect} it follows that to exclude 
the possibility of an LHV1 model
for the entangled Werner states (or for 
any other states admitting an LHV1
model)  POV's 
that are close to PV's in the
sense that they contain only commuting 
operators will not help.

Furthermore, we have extended Fine's 
result on the equivalence between the
existence of deterministic and stochastic 
LHV1 models to LCHVG
models. Besides the usefulness of this 
result for the construction of
models, we regard it as further evidence 
that our notion of a local causal
hidden variables model adequately captures 
the relevant physical ideas.

\section*{Acknowledgments}
We thank the referee for valuable 
comments and suggestions.
This work was supported in part  by
the DFG, by NSF Grant  No. DMS-9504556, 
and by the INFN. 

\begin{appendix}

\section{Equivalence between the existence 
of deterministic and stochastic hidden 
variables models}
\label{equisect}

A (deterministic) LCHV model $(\Omega, 
{\rm I\! P}, X)$ forms a degenerate 
stochastic LCHV model if one defines 
$\widetilde{\Omega} = \Omega$, 
$\widetilde{{\rm I\! P}} = {\rm I\! P}$, and sets
$  Q_{O^1 (t_1) ,\ldots ,O^n (t_n) } 
( o_{1} ,\ldots,o_{n} | \,
\widetilde\omega ) := 1$ if $X_{O^1 
(t_1) ,\ldots ,O^n
 (t_n)}(\widetilde\omega) = ( o _{1} ,
\ldots,o_{n})$ and  0 otherwise.
 Conversely, from a given stochastic 
model one can construct a
 (deterministic) LCHV model as follows: 
Let $\rho$ be a quantum state with
 a stochastic LCHV model $(\widetilde 
\Omega, \widetilde{{\rm I\! P}} , Q)$.  
For all $n\in {\rm I\! N}$,
 all sequences of local observables 
$A^k$ for the first subsystem and  all
 $a_k\in \sigma(A^k)$,\ $k=1\dots n$, 
and all $\widetilde \omega \in
 \widetilde \Omega$, there exist 
independent random variables
\[ \widehat X_{A^n,a_1,\dots ,a_{n-1},
\widetilde\omega} ^{A^1,\dots ,A^{n-1}} : 
\Omega _A \rightarrow \sigma(A^n) \] 
with distribution ${\rm I\! P}_A (\widehat
X_{A^n,a_1,\dots ,a_{n-1},\widetilde\omega} 
^{A^1,\dots ,A^{n-1}} = a_n
)=Q_{A^1,\dots ,A^n}(a_n |a_1,\dots , 
a_{n-1}, \widetilde \omega )$ on some
probability space $(\Omega _A, 
{\rm I\! P} _A)$. (The canonical choice is the huge
product space $\prod_\alpha {\rm I\! R}$ 
over all possible such choices $\alpha$,
equipped with the product measure 
$\prod_\alpha \mu_\alpha$ with
$\mu_\alpha$ the conditional probability
distribution $Q_{A^1,\dots
,A^n}(a_n |a_1,\dots , a_{n-1}, \widetilde 
\omega )$ corresponding to
$\alpha$, and with the random variables 
given by the corresponding
projections.) Similarly, for all $m\in 
{\rm I\! N}$, all sequences  of local
observables $B^k$ for the second subsystem 
and all $b_k\in \sigma(B^k)$,
$k=1\dots m$, and all $\widetilde \omega 
\in \widetilde \Omega$ define
independent random variables
\[ \widehat X_{B^m,b_1,\dots ,b_{m-1},
\widetilde\omega} ^{B^1,\dots ,B^{m-1}} : 
\Omega _B \rightarrow \sigma(B^m) \]
on some probability space $(\Omega_B, 
{\rm I\! P}_B)$ with the distribution ${\rm I\! P}_B 
(\widehat X_{B^m,b_1,\dots ,b_{m-1}, 
\widetilde\omega} ^{B^1,\dots ,B^{m-1}} 
= b_m )$ $ =$  $  Q_{B^1,\dots ,B^m} 
(b_m | b_1,\dots , b_{m-1}, \widetilde \omega )$.
Now put 
\[ \Omega = \widetilde\Omega \times \Omega 
_A\times \Omega_B, \quad {\rm I\! P} = 
\widetilde{\rm I\! P} \times {\rm I\! P}_A 
\times {\rm I\! P}_B \]
and define inductively all $X_{A^1,\dots , 
A^n}$ and  $X_{B^1,\dots , B^m}$ using 
(\ref{caus}): for $\omega = (\widetilde\omega, 
\omega_A, \omega_B)$
\[ X_{A^1} (\omega)= \widehat X_{A^1,
\widetilde\omega}(\omega_A), \quad X_{B^1} (\omega)= 
\widehat X_{B^1,\widetilde\omega}(\omega_B) , \]
\begin{eqnarray*}  X_{A^n}^{A^1,\dots ,A^{n-1}} 
(\omega) & = & \widehat 
X^{A^1,\dots ,A^{n-1}}_{A^n,X_{A^1,\dots , 
A^{n-1}}(\omega),\widetilde \omega} (\omega_A),\\
X_{B^m}^{B^1,\dots ,B^{m-1}} (\omega) & = & 
\widehat X^{B^1,\dots ,B^{m-1}}_{B^m,X_{B^1,
\dots , B^{m-1}}(\omega),\widetilde \omega} 
(\omega_B) . \end{eqnarray*}  
One easily sees that this defines an LCHV 
model for the quantum state $\rho$. (For example,
\begin{eqnarray*} & & {\rm I\! P} \bigl(X_{A^1,A^2} 
= (a_1,a_2)\bigr) \\
& & \quad =  \int_{\widetilde \Omega} 
d\widetilde {\rm I\! P} (\widetilde \omega)
\int_{\Omega_A} \!\!\!\!
d{\rm I\! P}_A (\omega_A) \int_{\Omega_B} \!\!\!\!
d{\rm I\! P}_B (\omega_B) \openone_{ \{ X_{A^1}=a_1, 
X_{A^2}^{A^1}= a_2 \} } \\
& & \quad =  \int_{\widetilde \Omega} d\widetilde 
{\rm I\! P} (\widetilde \omega)
\int_{\Omega_A} \!\!\!\!
d{\rm I\! P}_A (\omega_A) \int_{\Omega_B} \!\!\!\!
d{\rm I\! P}_B (\omega_B) \\
& & \hspace{4cm}\openone_{ \{ \widehat 
X_{A^1,\widetilde\omega} (\omega_A) =a_1, \widehat 
X_{A^2,a_1,\widetilde\omega }^{A^1} 
(\omega_A) = a_2 \} } \\
& & \quad =  \int_{\widetilde \Omega} d\widetilde 
{\rm I\! P} (\widetilde \omega)
{\rm I\! P}_A (\widehat X_{A^1,\widetilde\omega} 
=a_1,\widehat X_{A^2,a_1,
\widetilde\omega }^{A^1} = a_2) \\
& & \quad =  \int_{\widetilde \Omega} d\widetilde 
{\rm I\! P} (\widetilde \omega)
{\rm I\! P}_A (\widehat X_{A^1,\widetilde\omega} 
=a_1) {\rm I\! P}_A (\widehat X_{A^2,a_1,
\widetilde\omega }^{A^1} = a_2)  \\
& & \quad =  \int_{\widetilde \Omega} d\widetilde 
{\rm I\! P} (\widetilde \omega)
Q_{A^1}(a_1|\widetilde\omega) Q_{A^1,A^2}
(a_2|a_1, \widetilde\omega )\\
& & \quad = \int_{\widetilde \Omega} d\widetilde 
{\rm I\! P} (\widetilde \omega) 
Q_{A^1,A^2}(a_1, a_2| \widetilde\omega),
\end{eqnarray*}
where the independence of the random variables 
$\widehat X$ is used for the
4th equation.) 
\end{appendix}


\begin{references}

\bibitem{Bell}  J.S. Bell, {\it Speakable 
and Unspeakable in
Quantum Mechanics}\/  (Cambridge University 
Press, Cambridge, 1987).
\bibitem{EPR} A. Einstein, B. Podolsky, and 
N. Rosen, Phys. Rev. {\bf 47}, 777 (1935).
\bibitem{Aspect} A. Aspect, P. Grangier, and
G. Roger, Phys. Rev. Lett. {\bf 49}, 1804 (1982).
\bibitem{BohmEPR} D. Bohm, {\it Quantum Theory}\/ 
(Prentice-Hall, Englewood Cliffs,  1951).
\bibitem{Fine} A. Fine, Phys. Rev. Lett. 
{\bf 48}, 291 (1982).
\bibitem{Gisin} N. Gisin, Phys. Lett. A 
{\bf 145}, 201 (1991).
\bibitem{PopRohr} S. Popescu and D. Rohrlich, 
Phys. Lett. A {\bf 166}, 293 (1992).
\bibitem{Werner} R.F. Werner, Phys. Rev. 
{\bf A 40}, 4277 (1989).
\bibitem{Popescu} S. Popescu, Phys. Rev. Lett. 
{\bf 74}, 2619 (1995).
\bibitem{Gisin2} N. Gisin, Phys. Lett. A {\bf 210}, 
151 (1996).
\bibitem{MerminBiel} N.D. Mermin, {\it Notes 
for a lecture given in Bielefeld}, 
July 1995 (unpublished).
\bibitem{ZHHH} M. Zukowski, R. Horodecki, M. 
Horodecki, and P. Horodecki, {\it Generalized
measurements and local realism}, preprint 
(quant-ph/9608035).
\bibitem{nrao} M. Daumer, D. D\"urr, S. 
Goldstein, and N. Zangh\`\i, in {\it Erkenntnis}, 
Proceedings of the
International Conference on Probability, 
Dynamics and Causality, Luino,
Italy, 1995, edited by D. Costantini and
M.C. Gallavotti (to appear) (quant-ph/9601013).
\bibitem{Diplom} S. Teufel, Diplomarbeit, 
Universit\"at M\"unchen, 1996.
\bibitem{Peres} A. Peres, Phys. Rev. Lett. 
{\bf 77}, 1413 (1996).
\bibitem{Horodecki} M. Horodecki, P. Horodecki, and 
R. Horodecki, Phys. Lett. A {\bf 223}, 1 (1996).
\bibitem{Purif} C.H. Bennett, G. Brassard, S. Popescu, 
B. Schumacher, J. Smolin, and W.K. 
Wootters, Phys. Rev. Lett. {\bf 76}, 722 (1996).
\bibitem{Davies} E.B. Davies, {\em Quantum Theory of 
Open Systems} (Academic Press, 1976).
\bibitem{DDGZ} M. Daumer, D. D\"urr, S. Goldstein, and 
N. Zangh\`\i, {\it Bohmian mechanics and the
role of operators as observables in quantum theory}, 
in preparation.
\bibitem{CommGis} K. Berndl and  S. Teufel, 
Phys. Lett. A {\bf 224}, 314 (1997).
\end{references}
\end{document}